Growth and Characterization of Superconducting Bulk Crystal

[(SnSe)$_{1+\delta}$]$_m$(NbSe$_2$) Misfit Layer Compounds

Ryufa Shu[1], Masanori Nagao[1*], Chiaya Yamamoto[1], Keisuke Arimoto[1], Junji Yamanaka[1],

Yuki Maruyama[1], Satoshi Watauchi[1], and Isao Tanaka[1]

[1]*University of Yamanashi, 7-32 Miyamae, Kofu, Yamanashi 400-0021, Japan*

[*]Corresponding Author

Masanori Nagao

Postal address: University of Yamanashi, Center for Crystal Science and Technology

Miyamae 7-32, Kofu 400-0021, Japan

Telephone number: (+81)55-220-8610

Fax number: (+81)55-220-8270

E-mail address: mnagao@yamanashi.ac.jp

**Abstract**

$[(SnSe)_{1+\delta}]_m(NbSe_2)$ ($m$ = 1–6, 8, and 12) highly orientated crystals 1–2 mm in size and well-defined $c$-planes were successfully grown using CsCl/KCl flux, including the first growth of crystals with $m$ = 12. The stacked layers along the $c$ axis in the obtained crystals were directly observed by transmission electron microscopy as $m$ alternating layers of SnSe and single layers of $NbSe_2$. The superconducting transition temperature of the obtained $[(SnSe)_{1+\delta}]_m(NbSe_2)$ crystals decreased with an increase in the number of SnSe layers per unit cell. As the superconducting anisotropy parameters increase, a significant increase is observed between $m$ = 4 and 5. This indicates that the superconducting dimensionality becomes more two-dimensional with an increasing $m$.

**Main text**

## 1. Introduction

  Misfit layer compounds (MLCs) are composed of alternating layers of $MX$ ($M$ = Sn, Pb,

Sb, Bi, or a lanthanide; $X$ = S, Se, or Te) with a rock salt structure, and $TX_2$ ($T$ = Ti, V,

Cr, Nb, or Ta) layers of transition metal dichalcogenides (TMDCs) held on a sublattice

stacked along the $c$ axis. MLCs undergo atomic positional modulation due to their

incommensurate layered structures along the $b$ axis, resulting in unique physical

properties [1]. Furthermore, MLCs nanotubes with one-dimensional structures have also

been investigated [2]. The general formula for MLCs can be expressed as $[(MX)_{1+\delta}]_m(TX_2)$

($\delta$ = 0.08–0.28, $m$ = positive integer), where the incommensurate factor is indicated by

$1+\delta$ [3–6]. TMDCs can exhibit two-dimensional (2D) metallic, semiconductor, or

superconductor behaviors owing to various element combinations. Examples of application propositions include anode materials for sodium-ion batteries [7],

photocatalysts for water splitting [8], platform materials for heavily doped 2D

dichalcogenides [9], and emergent superconductivity using field-effect transistors [10].

Because of their individualistic cleavage feature, MLCs can be exfoliated and restacked,

and the resulting improvement of their thermoelectric performance has also been reported

[11]. They have received considerable attention in a wide range of research fields and

have many untapped properties [12]. MLCs often exhibit superconductivity, including those comprising Nb, Ta, and Ti dichalcogenides with a rock salt structure and Sn, Pb, Bi, and La monochalcogenide layers [13–23]. In particular, a superconducting transition temperature ($T_c$) of approximately 7.2 K and additional advantageous physical properties such as charge density waves were observed in NbSe$_2$ [24,25]. NbSe$_2$ can be easily exfoliated owing to the van der Waals forces between the layers. The $T_c$ of the monolayered NbSe$_2$ is approximately 3.1 K, which is lower than that of the bulk. Moreover, the physical properties of monolayer and multilayer NbSe$_2$ are distinct [26]. Exceptional thermoelectric efficiency has been observed in SnSe, a monochalcogenide with a rock salt structure. It has an ultralow thermal conductivity and high thermoelectric figure of merit [27].

In this paper, we focus on the $[(SnSe)_{1+\delta}]_m(NbSe_2)$ MLCs that exhibit unique physical properties and superconductivity. Crystalline samples are typically grown via chemical vapor transport (CVT) [28,29] and physical vapor deposition (PVD) [30,31]. However, these methods require meticulous control of the growth conditions and exhibit slow growth rates. Additionally, layered chalcogenide single crystals were grown using the flux method [32–35]. The flux method uses simple equipment and achieves higher growth rates than those of the CVT and PVD methods. Herein, we have successfully grown [(SnSe)$_{1+\delta}$]$_m$(NbSe$_2$) crystals using the flux method. The growth conditions of [(SnSe)$_{1+\delta}$]$_m$(NbSe$_2$) crystals with various $m$ values (number of SnSe layers) were optimized, including the growth temperature and nominal composition. The superconducting properties were evaluated by increasing the number of SnSe layers ($m$) in a unit cell. The resulting superconducting properties were compared to those of a previous report on [(SnSe)$_{1+\delta}$]$_m$(NbSe$_2$) thin films [23]. The differences between the superconducting properties of the thin films and bulk crystals were investigated.

## 2. Experimental

[(SnSe)$_{1+\delta}$]$_m$(NbSe$_2$) crystals were grown using a CsCl/KCl flux method. The raw materials Nb (99.9%, Powder: High Purity Chemicals), Sn (99.9%, Drops: Kanto Chemicals), and Se (99.99%, Powder: Rare Metallic) were weighed using a nominal composition of NbSn$_2$Se$_4$. The molar ratio of the flux was CsCl (99.8%, Powder: Kanto Chemicals):KCl (99.5%, Powder: Kanto Chemicals) = 5:3. A mixture of the raw materials (0.8 g) and flux (5.0 g) was ground in a mortar and sealed in an evacuated quartz tube (Approx. 10 Pa). The quartz tube was heated to 800 °C for 10 h and subsequently cooled to 600 °C at a rate of 1 °C/h. The temperature was then decreased to 400 °C at a rate of 100 °C/h and maintained at 400 °C for 100 h. The sample was then allowed to cool to less than 30 °C in the furnace. The heat-treated quartz tube was opened to air, and the obtained products were washed and filtered using distilled water to remove the CsCl/KCl flux and hydrosoluble impurities.

The microstructures of the obtained crystals were observed using scanning electron microscopy (SEM) (Hitachi High-Technologies, TM3030), and the compositional ratio was estimated using energy-dispersive X-ray spectrometry (EDS) (Bruker, Quantax 70). The measured compositional values were normalized using Nb = 1 and the Sn and Se compositions were determined. The crystals were identified and oriented using X-ray diffraction (XRD, Rigaku MultiFlex) with CuK$\alpha$ radiation. The $m$ values of the crystals obtained were determined. Samples were prepared for transmission electron microscopy (TEM) using a Ga$^+$ ion type focused ion beam (HITACHI FB-2100A) with an acceleration voltage of 40 kV after depositing the protection layers (C, Pt–Pd, and W). TEM observations were performed using a field-emission transmission electron microscope (FE-TEM) (FEI, Tecnai Osiris) at an acceleration voltage of 200 kV.

The Hall measurements and resistivity–temperature ($\rho$–$T$) characteristics of the obtained crystals were determined using the standard four-probe method in constant current ($J$) mode with a physical property measurement system (PPMS; Quantum Design

DynaCool). The obtained sample was placed on an MgO single-crystal substrate, and the electrical terminals were made of silver paste (DuPont; 4922N).

The $\rho$–$T$ characteristics below 1.8 K were measured using the adiabatic demagnetization refrigerator (ADR) option of the PPMS. The magnetic field applied to operate the ADR was 3 T at 1.9 K. Subsequently, the magnetic field was removed. Consequently, the temperature of the sample decreased to approximately 0.13 K. The measurement of the $\rho$–$T$ characteristics was commenced at the lowest temperature (approximately 0.13 K), which was spontaneously increased to 15 K. The transition temperature corresponding to the onset of superconductivity ($T_c^{onset}$) is defined as the temperature at which a deviation from linear behavior is observed in the normal conducting state $\rho$–$T$ characteristics. Zero resistivity ($T_c^{zero}$) is defined as the temperature at which the resistivity is below approximately 1.0 $\mu\Omega$ cm in the $\rho$–$T$ characteristic.

The upper critical field ($H_{c2}$) was estimated from the field dependence of $T_c^{onset}$ under the magnetic field ($H$). The superconducting anisotropic parameter ($\gamma_s$) was evaluated using two methods. The first is the ratio of $H_{c2}$ parallel to the $c$-plane ($H$//$c$-plane) to $H_{c2}$ parallel to the $c$ axis ($H$//$c$-axis). In the second method, the angular ($\theta$) dependence of resistivity ($\rho$) in the flux liquid state was measured under various magnetic fields ($H$), and the superconducting anisotropic parameter ($\gamma_s$) calculated using the effective mass model

[36].

## 3. Results and discussion

The [(SnSe)$_{1+\delta}$]$_m$(NbSe$_2$) crystals with various $m$ values were successfully obtained.

Figure 1 shows a SEM image of the obtained [(SnSe)$_{1+\delta}$]$_m$(NbSe$_2$) crystal with $m = 6$. The

crystal could be easily cleaved and had a plate-like shape with a size and thickness of 1–

2 mm and approximately 50 μm, respectively. Furthermore, the crystal surface exhibited metallic luster and moiré-like patterns.

Figure 2 shows the XRD patterns of the obtained [(SnSe)$_{1+\delta}$]$_m$(NbSe$_2$) crystals, which

exhibited well-developed planes. The presence of only the 00$l$ diffraction peaks in Figure

2 (b) indicates that the $c$-plane was well-developed, which corresponded to

[(SnSe)$_{1+\delta}$]$_m$(NbSe$_2$) structures with various $m$ values [20,21,30]. The obtained

[(SnSe)$_{1+\delta}$]$_m$(NbSe$_2$) crystals with various $m$ values were confirmed to be highly

orientated, which enabled the estimation of the $m$ values. The XRD pattern labeled "$m$ =

12" has eight distinct diffraction peaks between those labeled "00$\underline{13}$" and "00$\underline{26}$". The distance of the 2$\theta$ value between the "00$\underline{19}$" and "00$\underline{24}$" peaks corresponded to the distance of including four diffraction peaks ("00$\underline{20}$" to "00$\underline{23}$" peaks). Therefore, the number of diffraction peaks between "00$\underline{13}$" and "00$\underline{26}$" was estimated to be twelve, corresponding to $m = 12$. The $c$-axis lattice constant was 7.51 nm using the assigned lattice indices of the diffraction peaks. This corresponds to $m = 12$, as shown in Figure 3. $[(SnSe)_{1+\delta}]_m(NbSe_2)$ crystals with $m = 2$–10 were previously reported as thin films [30]. Thus, we not only successfully obtained $[(SnSe)_{1+\delta}]_m(NbSe_2)$ bulk crystals with $m = 2$–6, and 8 but also grew $[(SnSe)_{1+\delta}]_m(NbSe_2)$ crystals with $m = 12$ for the first time. The $c$-axis lattice constants were determined from the XRD patterns shown in Figure 2. Figure 3 shows the $c$-axis lattice constants as a function of $m$ (number of SnSe layers) for the obtained $[(SnSe)_{1+\delta}]_m(NbSe_2)$ crystals. These $c$-axis lattice constants for each $m$ are consistent with a previous report on thin films [30]. Linear extrapolation indicated that the $c$-axis lattice constant approached 0.653 nm at $m = 0$, which corresponded to the thickness of a single NbSe$_2$ layer [37]. However, the gradient of this extrapolated line was 0.572 nm, which is similar to the $c$-axis lattice constant of SnSe [38]. Table I shows the estimated $m$ values and analytical compositions of the obtained $[(SnSe)_{1+\delta}]_m(NbSe_2)$ crystals. The evaluated analytical compositions closely matched the estimated $m$ values. However, deviations from the estimated $m$ values were observed. In particular, the error range increases at high $m$ values. This suggests the intergrowth of other $m$-value layers in the obtained crystals. Although $[(SnSe)_{1+\delta}]_m(NbSe_2)$ is an MLC, the lattice constants of the subunit cell have to be determined for the estimation of the precise compositions, i.e., it is necessary to estimate the $\delta$ values [1].

Figure 4 shows the cross-sectional TEM lattice images of the (a) $m = 1$ and (b) $m = 3$ crystals. The difference between $m = 1$ and 3 corresponds to the number of SnSe layers. The distance between $NbSe_2$ layers in $m = 1$ and 3 were 1.2 and 2.4 nm, respectively. These values agree well with those shown in Fig. 3 and correspond to the lengths of $c$-axis stacking layers ($m = 1$: $SnSe + NbSe_2$, $m = 3$: $SnSe \times 3 + NbSe_2$).

Based on the Hall measurements, the obtained $[(SnSe)_{1+\delta}]_m(NbSe_2)$ crystals are p-type conductors with holes as the dominant carriers. Figure 5 shows dependence of the number of SnSe layers ($m$) on the carrier concentrations estimated from the Hall measurements for the grown $[(SnSe)_{1+\delta}]_m(NbSe_2)$ crystals. The carrier concentrations were in the range of $10^{21}$ cm$^{-3}$, with a small dependence on temperature, and decreased with increasing $m$. These behaviors are almost consistent with those of thin-film samples [30]. In a previous report [30], the carrier concentration decreased significantly between $m = 5$ and 6. However, this was observed between $m = 2$ and 3 in the current result.

Figure 6 shows the resistivity-temperature ($\rho$–$T$) characteristics of the obtained $[(SnSe)_{1+\delta}]_m(NbSe_2)$ crystals. As shown in Figure 6 (b), all the obtained $[(SnSe)_{1+\delta}]_m(NbSe_2)$ crystals exhibit zero resistivity ($T_c^{zero}$), which indicated the presence of superconductivity. The resistivities of the normal state above $T_c^{\text{zero}}$ for $m = 8$ and $12$ are significantly higher than those of the other samples. However, these values did not correlate with the number of SnSe layers ($m$). The dependence of $T_c^{\text{zero}}$ and $T_c^{\text{onset}}$ on $m$ is shown in Figure 7; both $T_c^{\text{zero}}$ and $T_c^{\text{onset}}$ decreased with increasing $m$. The observed $T_c^{\text{zero}}$ value was higher than that previously reported for thin-film experiment [23]. Furthermore, $T_c^{\text{zero}}$ was observed for the first time for $[(\text{SnSe})_{1+\delta}]_m(\text{NbSe}_2)$ compounds with $m \geq 8$. However, $T_c^{\text{zero}}$ decreases moderately with an increase in $m$. We assumed the possibility of the intergrowth of a few other $m$-value layers in these crystals.

We hypothesized that the superconductivity in $[(\text{SnSe})_{1+\delta}]_m(\text{NbSe}_2)$ originated from the $\text{NbSe}_2$ layers. As the number of SnSe layers ($m$) in $[(\text{SnSe})_{1+\delta}]_m(\text{NbSe}_2)$ compounds increased, the superconducting coupling between the $\text{NbSe}_2$ layers weakened [30]. Therefore, the $\text{NbSe}_2$ layers may exhibit a 2D behavior. In a previous report [39], the superconducting transition temperature of $\text{NbSe}_2$ was drastically reduced by decreasing the number of layers to less than several unit layers. The behavior of the $\text{NbSe}_2$ superconductor was similar, although the superconducting transition temperature was not fully consistent. Thus, the $\text{NbSe}_2$ layers may be interacting with the SnSe layers in $[(\text{SnSe})_{1+\delta}]_m(\text{NbSe}_2)$ superconductors.

Figure 8 shows the resistivity-temperature ($\rho$–$T$) characteristics under the magnetic field ($H$) parallel to the (a) $c$ plane (0.1–9.0 T) and (b) $c$ axis (0.1–1.5 T) close to the superconducting transition temperature for $[(SnSe)_{1+\delta}]_3(NbSe_2)$ ($m = 3$). The suppression of the superconductivity was more significant under a magnetic field applied parallel to the $c$ axis than that applied parallel to the $c$-plane. This result suggested that $[(SnSe)_{1+\delta}]_m(NbSe_2)$ exhibited high superconducting anisotropy. This can be attributed to weak coupling between the $NbSe_2$ interlayer in the $c$-axis and $c$-plane directions. The influence of the magnetic field ($H$) applied parallel to the $c$ plane ($H//c$-plane) and $c$ axis ($H//c$-axis) on $T_c^{onset}$ is plotted in Figure 9. The linear extrapolation to $T_c^{onset} = 0$ K for $H//c$-plane and $H//c$-axis are 7.7 and 1.0 T, respectively. The upper critical fields $H_{c2}^{//c\text{-}plane}$ and $H_{c2}^{//c\text{-}plane}$ were estimated to be less than 7.7 and 1.0 T, respectively. Thus, the upper critical fields $H_{c2}^{//c\text{-}plane}$ and $H_{c2}^{//c\text{-}plane}$ were evaluated to be 6.5 and 0.9 T, respectively, by the Ginzburg–Landau equation [40]. The superconducting anisotropic parameter ($\gamma_s$) was evaluated from the ratio of the upper critical field ($H_{c2}$) using the following equation:

$$\gamma_s = H_{c2}^{//c\text{-}plane} / H_{c2}^{//c\text{-}axis} = \xi_{c\text{-}plane} / \xi_{c\text{-}axis} \qquad \text{(Eq. 1)}$$

where $\xi$ is the coherence length. The superconducting anisotropic parameter ($\gamma_s$) for $[(SnSe)_{1+\delta}]_3(NbSe_2)$ ($m = 3$) was calculated to be 7.7 and 7.2 by the linear extrapolation and the Ginzburg–Landau equation, respectively. We calculated $\gamma_s$ using the effective mass model [36]. The angular ($\theta$) dependence of resistivity ($\rho$) is measured at various magnetic fields ($H$) in the flux liquid state to estimate $\gamma_s$ [41,42]. The reduced field ($H_{red}$) was calculated using the following equation:

$$H_{red} = H(\sin^2\theta + \gamma_s^{-2}\cos^2\theta)^{1/2} \tag{Eq. 2}$$

where $\theta$ is the angle between the $c$-plane and the magnetic field [36]. The $\gamma_s$ value was estimated from the best scaling of the $\rho$–$H_{red}$ relations. Figure 10 shows the $\theta$ dependence of $\rho$ at various magnetic fields ($H = 0.1$–$2.0$ T) in the flux liquid state (directly above $T_c^{zero}$) for [(SnSe)$_{1+\delta}$]$_3$(NbSe$_2$) ($m = 3$) crystals. The $\rho$–$\theta$ curve exhibited two-fold symmetry. The resistivity originated from vortex flow, which increased with an increase in the strength of the applied field $H$. This phenomenon exhibited that the vortex flow velocity was increased with applied field $H$. Figure 11 shows the $\rho$–$H_{red}$ scaling obtained from the $\rho$–$\theta$ curves in Figure 10 using Eq. 2. The scaling was performed by taking $\gamma_s = 12$. The $\gamma_s$ of the other [(SnSe)$_{1+\delta}$]$_m$(NbSe$_2$) crystals were evaluated in the same way, which will be shown in Figure 12. The evaluated superconducting anisotropy parameter $\gamma_s$ and previously reported values for [(SnSe)$_{1+\delta}$]$_m$(NbSe$_2$) superconductors [23] are compared in Figure 12. The evaluated $\gamma_s$ between upper critical fields ($H_{c2}$) and the effective mass model exhibited a similar trend. The superconducting anisotropy increases with an increased number of SnSe layers ($m$) for $[(SnSe)_{1+\delta}]_m(NbSe_2)$. The previously reported $\gamma_s$ values for thin films [23] are comparable in the range of $m = 1$–3. However, the opposite trend was observed for the bulk samples and thin films. One possible explanation for this discrepancy is the influence of surface effects on the thin film; however, the reason for this requires further investigation. The $\gamma_s$ values of $[(SnSe)_{1+\delta}]_m(NbSe_2)$ crystals drastically increased between $m = 4$ and 5. This suggested that the superconductivity in $[(SnSe)_{1+\delta}]_m(NbSe_2)$ transitioned from three-dimension to two-dimension-like behavior.

The increase in the number of SnSe layers ($m$) enhances the localization of $NbSe_2$ layers in $[(SnSe)_{1+\delta}]_m(NbSe_2)$. 2D behavior was exhibited in the $NbSe_2$ layers, which are superconducting layers. Superconductivity in $[(SnSe)_{1+\delta}]_m(NbSe_2)$ with increasing SnSe layers may be affected by both the dimensionality of $NbSe_2$ superconducting layers and the interaction from the SnSe layers. Consequently, the superconducting phenomena of $[(SnSe)_{1+\delta}]_m(NbSe_2)$ are expected to have a complex behavior.

## 4. Conclusion

$[(SnSe)_{1+\delta}]_m(NbSe_2)$ ($m = 1$–6, 8, and 12) highly orientated crystals were successfully grown using a CsCl/KCl flux, which was confirmed by XRD patterns and the derived $c$- axis lattice parameters. Notably, the growth of $[(SnSe)_{1+\delta}]_m(NbSe_2)$ crystals with $m = 12$ is reported for the first time. The differences in the stacking layers along the $c$ axis between $m = 1$ and 3 are directly observed in the TEM images. The superconducting transition temperature ($T_c$) decreases with increasing $m$ (number of SnSe layers). However, the change in $T_c$ in the range of $m = 2$–12 was moderate compared to previous thin-film data. This may be due to the intergrowth of a few other $m$-value layers. The superconducting anisotropic parameters ($\gamma_s$) estimated using upper critical fields ($H_{c2}$) and the effective mass model yielded similar values. These values increase with the number of SnSe layers ($m$) and increase drastically between $m = 4$ and 5.

**Acknowledgments**

This work was partially supported by JSPS KAKENHI (Grant-in-Aid for Scientific Research (B) and (C): Grant Number 21H02022, 19K05248, and 23K03358, Grant-in-Aid for Challenging Exploratory Research: Grant Number 21K18834). We would like to thank Editage (www.editage.jp) for English language editing.

Table I. Analytical compositional ratio of the grown $[(SnSe)_{1+\delta}]_m(NbSe_2)$ crystals.

Values are normalized by setting Nb = 1.

| Estimated *m* value | Nb | Sn | Se |
|---|---|---|---|
| 1 | 1 | 1.15 ± 0.07 | 3.06 ± 0.33 |
| 2 | 1 | 2.28 ± 0.09 | 4.24 ± 0.05 |
| 3 | 1 | 3.85 ± 0.25 | 5.91 ± 0.49 |
| 4 | 1 | 4.40 ± 0.56 | 6.59 ± 0.51 |
| 5 | 1 | 5.88 ± 0.76 | 9.31 ± 1.09 |
| 6 | 1 | 6.53 ± 0.52 | 7.97 ± 1.08 |
| 8 | 1 | 8.98 ± 0.93 | 10.6 ± 1.0 |
| 12 | 1 | 10.8 ± 1.8 | 12.4 ± 2.0 |

**Figure captions**

Figure 1. Typical SEM image of a $[(SnSe)_{1+\delta}]_6(NbSe_2)$ ($m = 6$) crystal.

Figure 2. (a) XRD patterns of a well-developed plane of the grown $[(SnSe)_{1+\delta}]_m(NbSe_2)$ crystals. (b) Magnified XRD patterns of the 5–32° region.

Figure 3. Dependence of the $c$-axis lattice parameter for the grown $[(SnSe)_{1+\delta}]_m(NbSe_2)$ crystals on the number of (SnSe) layers ($m$).

Figure 4. Cross-sectional TEM lattice images of (a) $[(SnSe)_{1+\delta}](NbSe_2)$ ($m = 1$) and (b) $[(SnSe)_{1+\delta}]_3(NbSe_2)$ ($m = 3$) crystals. Inset: magnified view of the stacking structure.

Figure 5. Dependence on the carrier concentrations estimated from the Hall measurements for the number of SnSe layers ($m$) of grown $[(SnSe)_{1+\delta}]_m(NbSe_2)$ crystals.

Figure 6. (a) Temperature dependence of the electrical resistivity of the grown $[(SnSe)_{1+\delta}]_m(NbSe_2)$ crystals. (b) Magnified plot of the superconducting transition.

Figure 7. Dependence of the superconducting transition temperature with (a) zero resistivity ($T_c^{zero}$) and (b) onset of superconductivity ($T_c^{onset}$) on the number of SnSe layers ($m$) for the grown $[(SnSe)_{1+\delta}]_m(NbSe_2)$ crystals. Inset: schematic of the crystal structures partially derived from reference [30].

Figure 8. Temperature dependence of resistivity for the grown $[(SnSe)_{1+\delta}]_3(NbSe_2)$ ($m = 3$) crystal under magnetic fields of 0.1–9.0 T parallel to the (a) $c$ plane and (b) $c$ axis.

Figure 9. Field dependences of $T_c^{onset}$ under magnetic fields ($H$) parallel to the $c$ plane ($H//c$-plane) and $c$ axis ($H//c$-axis) from the data of Figure 7. The solid and dashed lines are the linear extrapolation and Ginzburg–Landau equation fits, respectively, to the data.

Figure 10. The angular ($\theta$) dependence of resistivity ($\rho$) in flux liquid state at various magnetic fields for the $[(SnSe)_{1+\delta}]_3(NbSe_2)$ ($m = 3$) crystal.

Figure 11. Reduced magnetic field $H_{red}$ dependence of resistivity ($\rho$) scaled by Eq. 2: $H_{red} = H(\sin^2\theta + \gamma_s^{-2}\cos^2\theta)^{1/2}$ using the data from Figure 9.

Figure 12. Dependence on the superconducting anisotropic parameters ($\gamma_s$) calculated from the upper critical field ($H_{c2}$) ratio and the effective mass model for the number of SnSe layers ($m$) of grown $[(SnSe)_{1+\delta}]_m(NbSe_2)$ crystals.

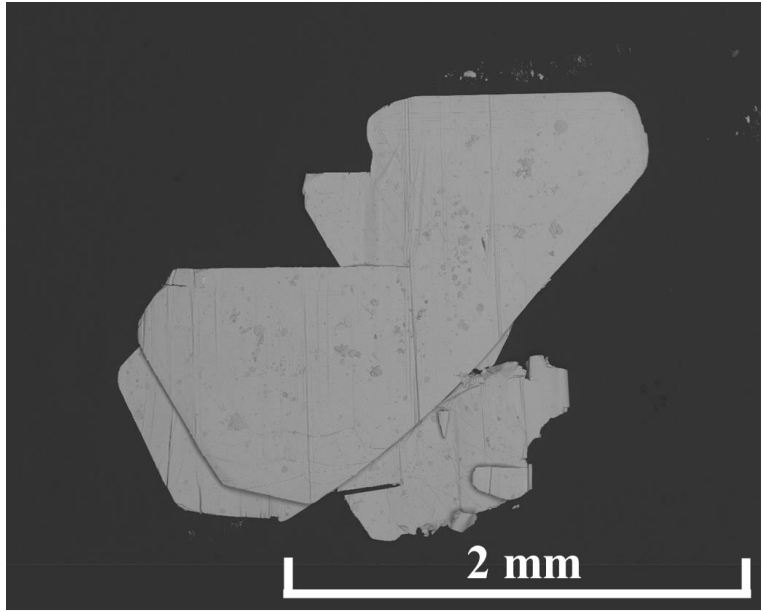

**Figure 1**

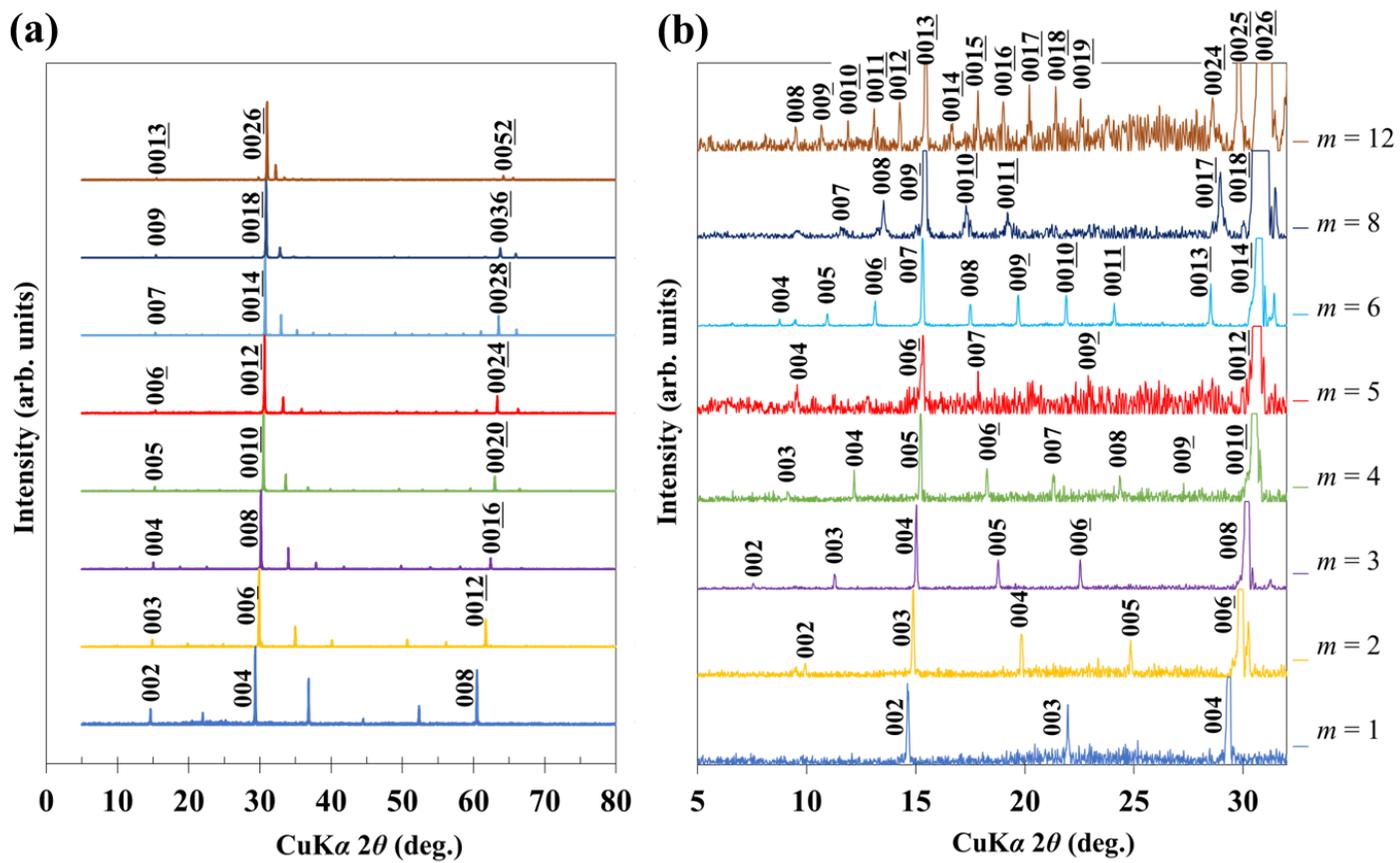

**Figure 2**

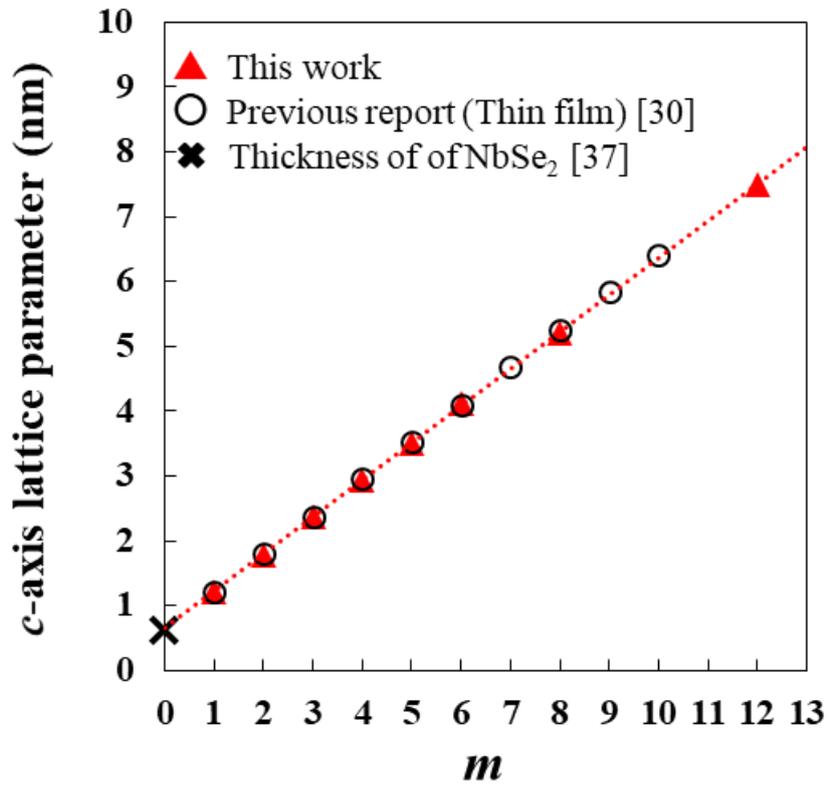

**Figure 3**

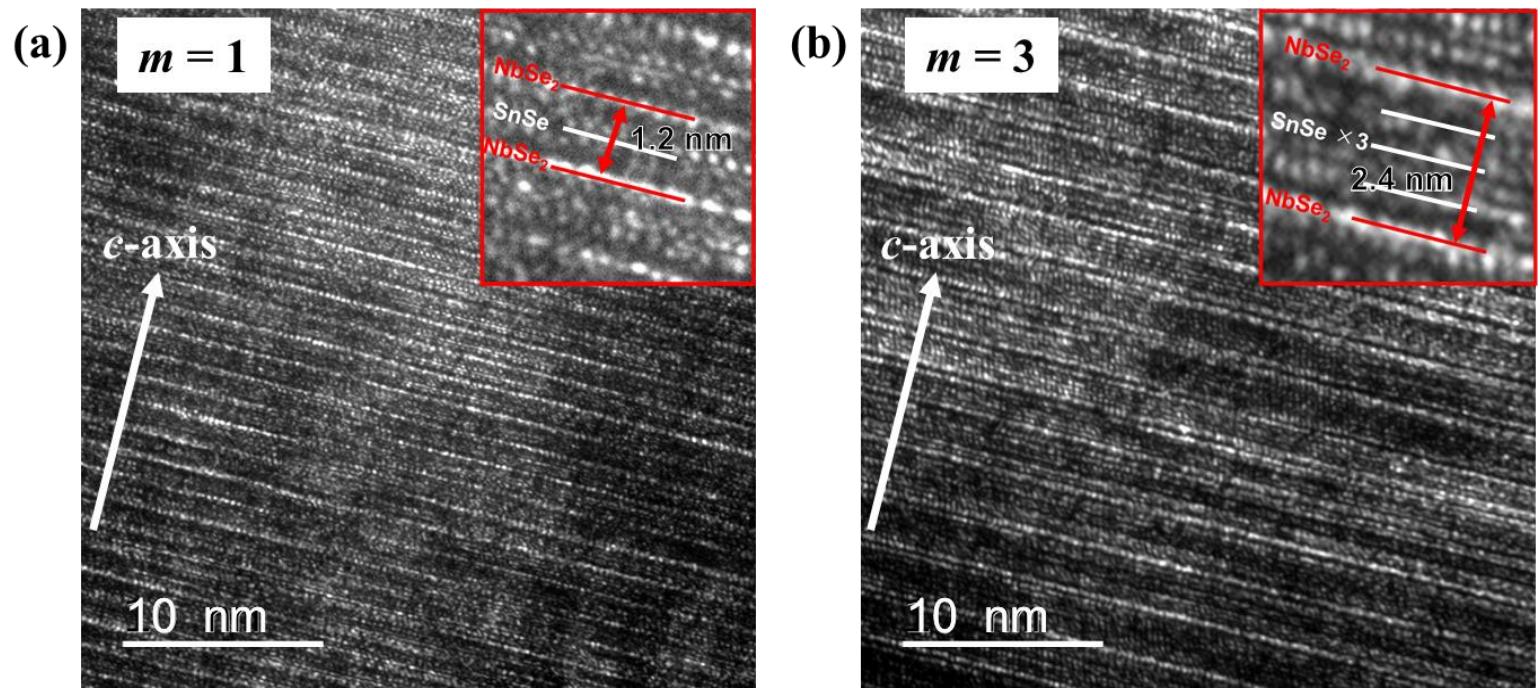

**Figure 4**

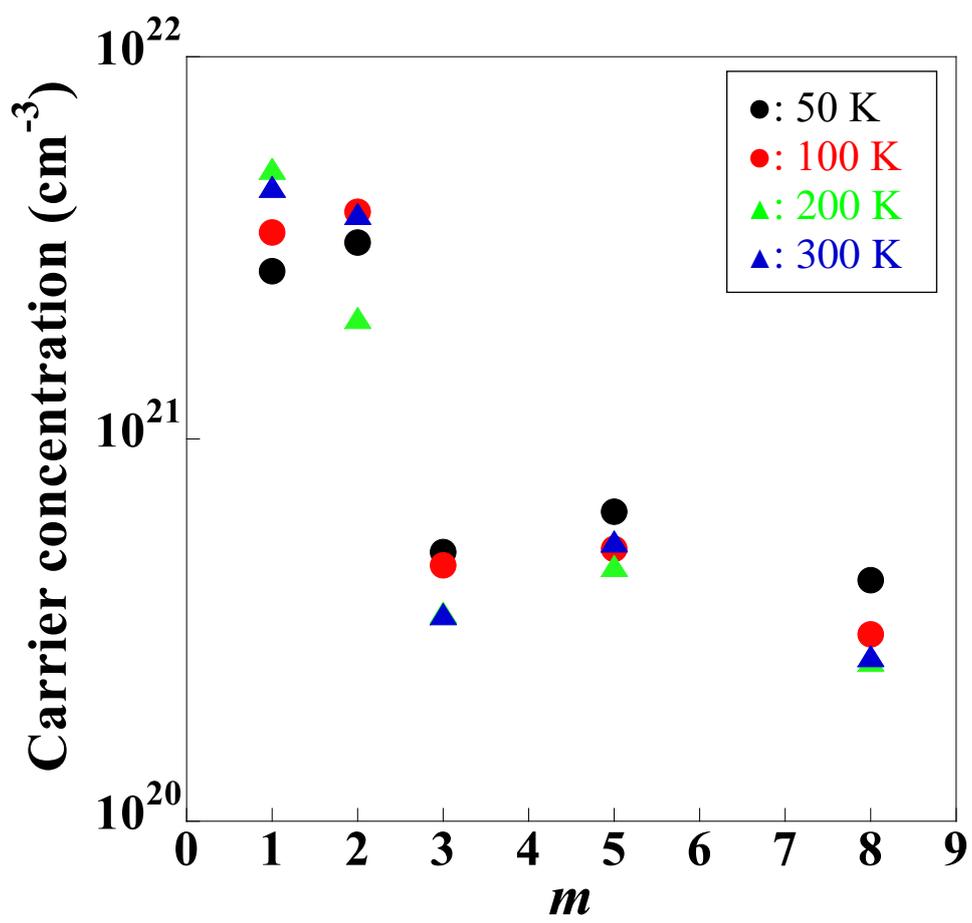

**Figure 5**

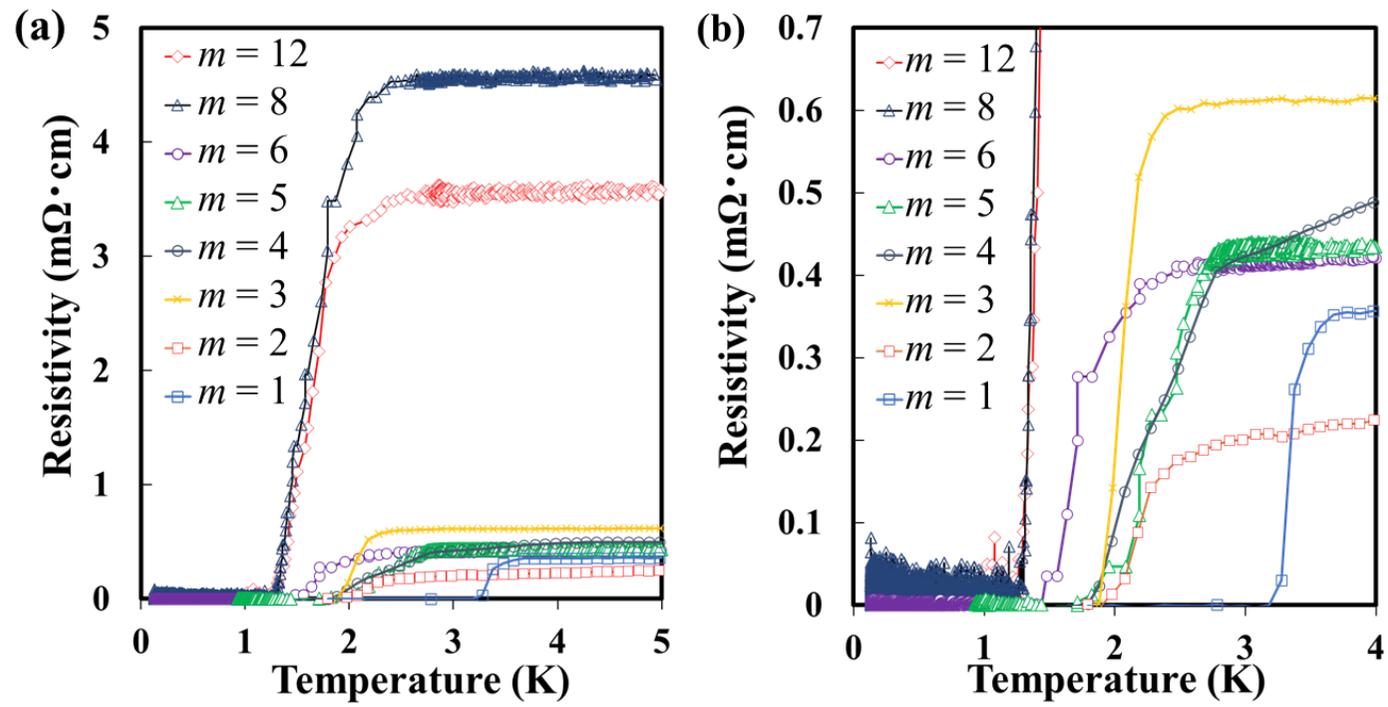

Figure 6

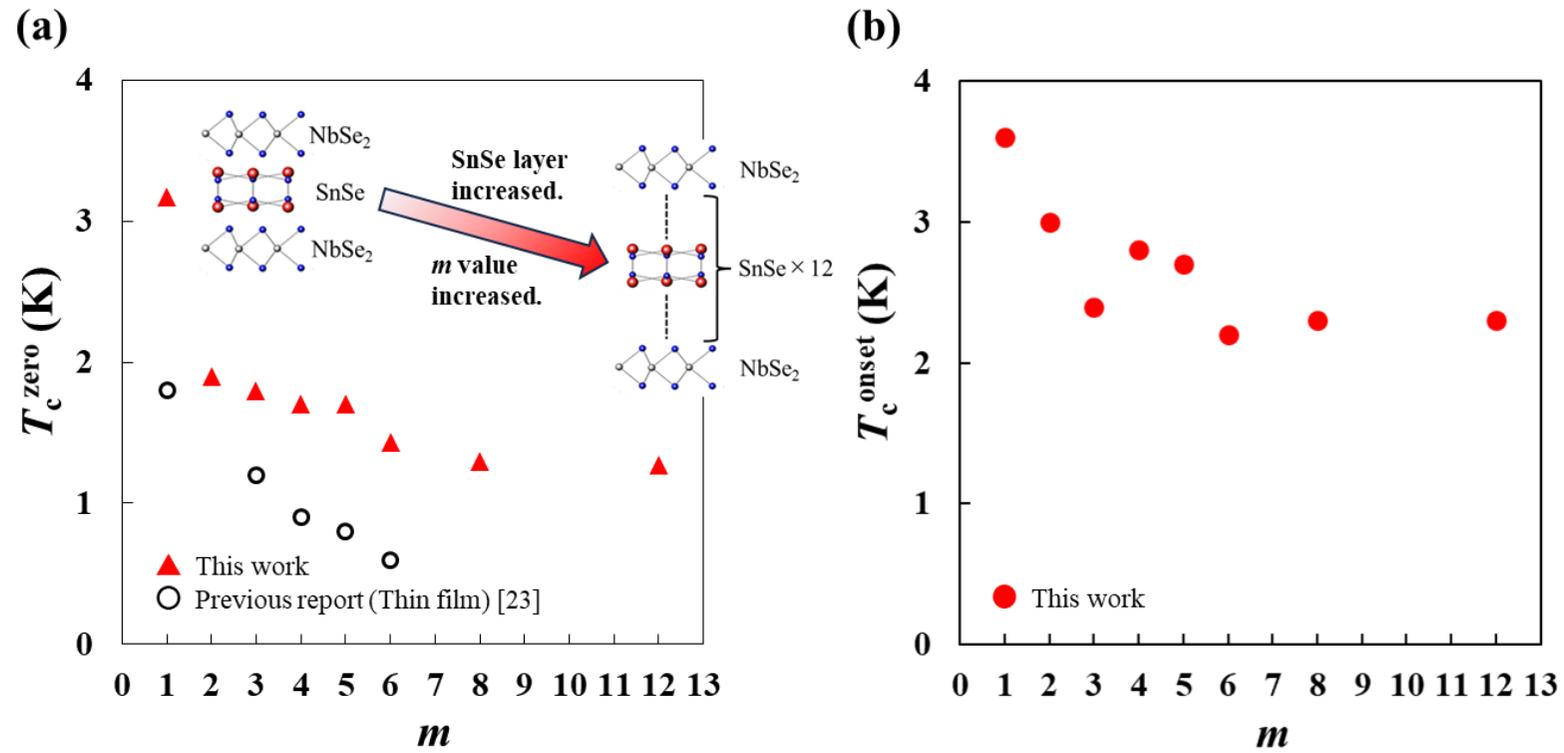

**Figure 7**

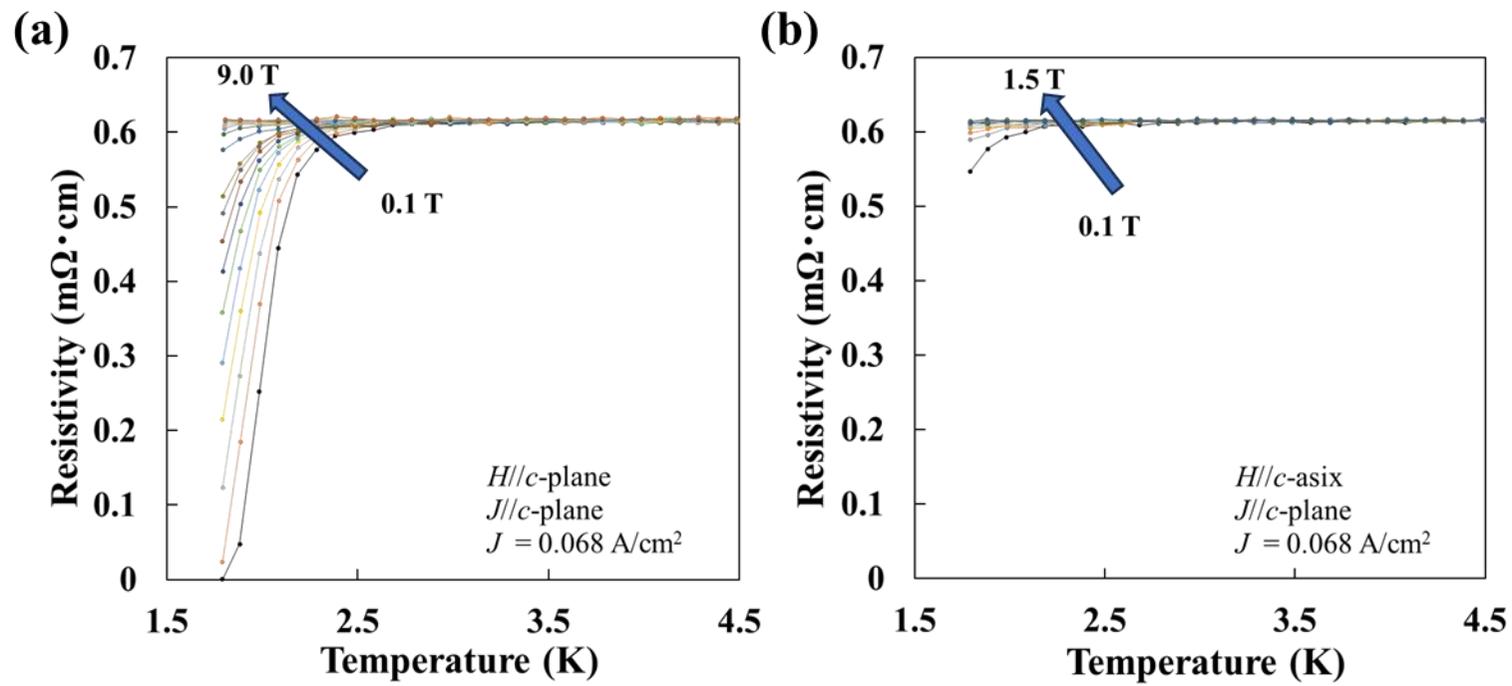

Figure 8

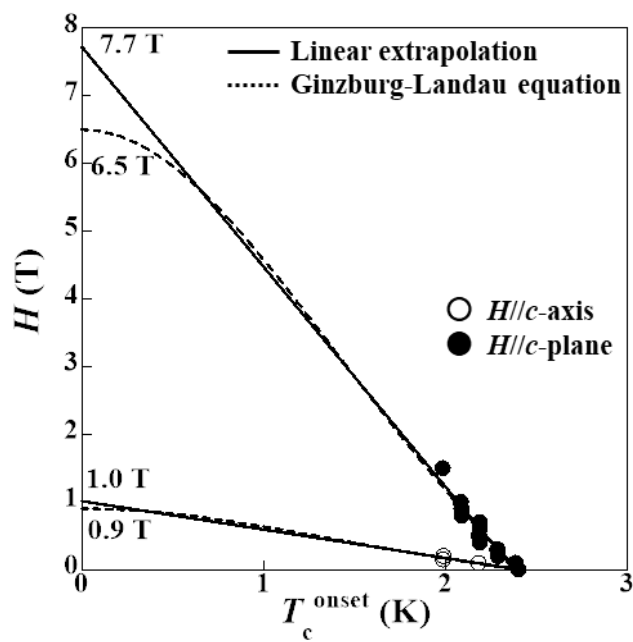

**Figure 9**

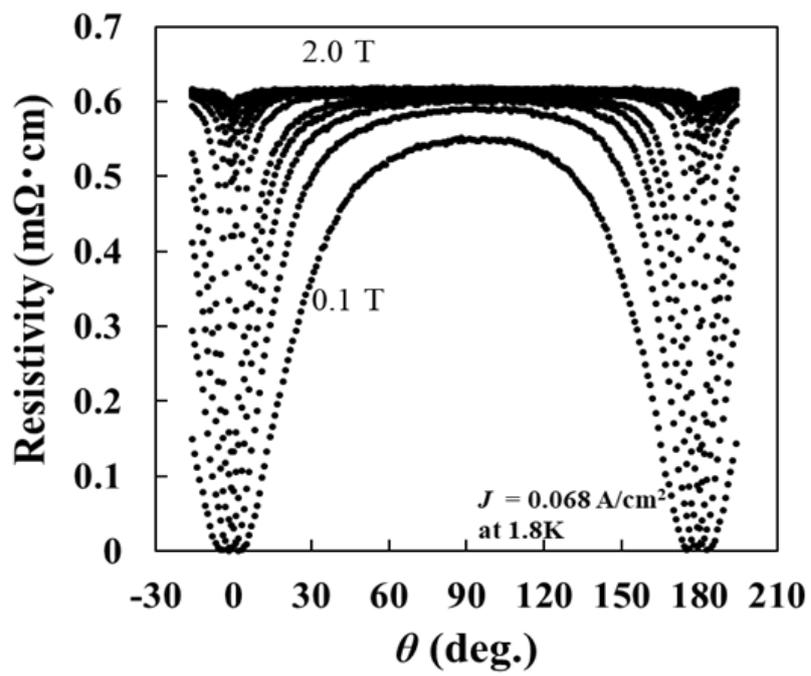

**Figure 10**

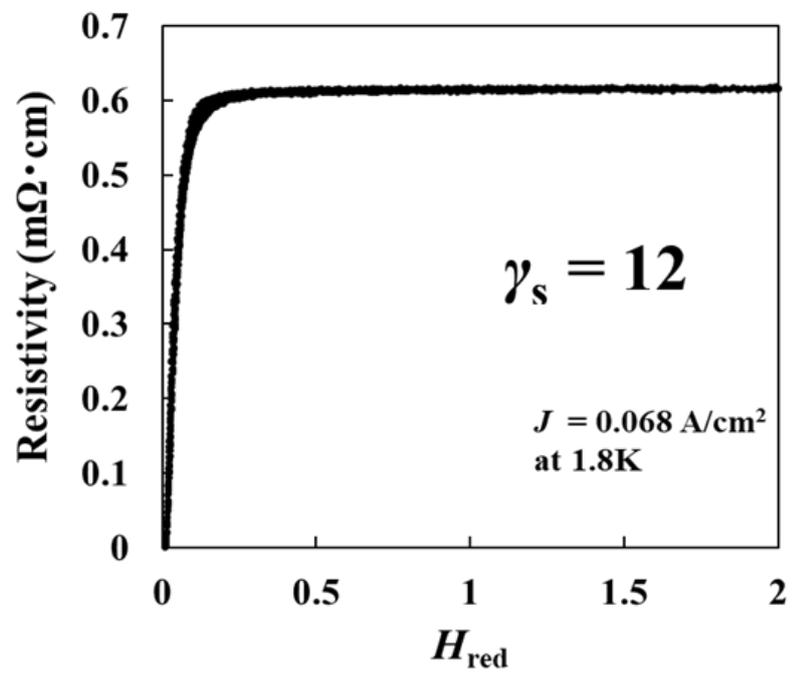

**Figure 11**

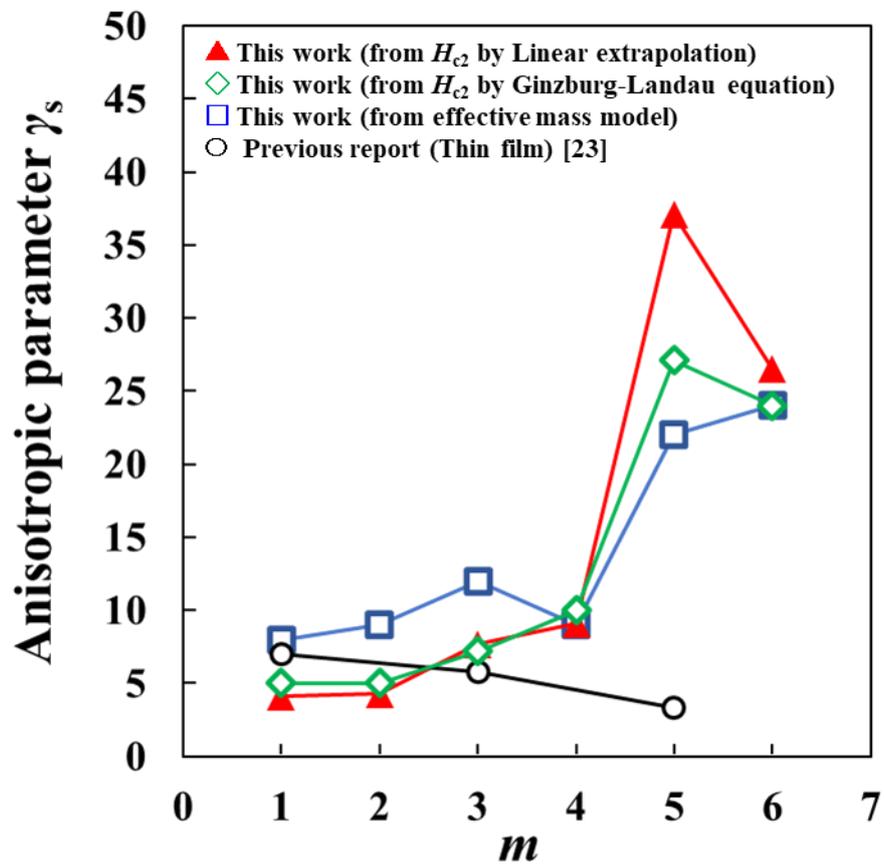

**Figure 12**